\documentclass[conference]{IEEEtran}
\IEEEoverridecommandlockouts
\usepackage{cite}
\usepackage{amsmath,amssymb,amsfonts}
\usepackage{algorithmic}
\usepackage{graphicx}
\usepackage{textcomp}
\usepackage{booktabs, multirow, array}
\usepackage{xcolor}
\usepackage{hyperref}
\def\BibTeX{{\rm B\kern-.05em{\sc i\kern-.025em b}\kern-.08em
    T\kern-.1667em\lower.7ex\hbox{E}\kern-.125emX}}
\begin{document}



\title{Feasibility Study of VLC-Based Collective Perception for Vehicular Communication}




\author{\IEEEauthorblockN{Kosuke NAKANO, Shan LU, Takaya YAMAZATO }
\IEEEauthorblockA{\textit{Dept. of Information and Communication Engineering},
\textit{Nagoya University}\\
Nagoya 464-8063, Japan\\
{knakano,lu,yamazato}@yamazato.nuee.nagoya-u.ac.jp}}

\maketitle

\begin{abstract}
This study explores the use of Visible Light Communication (VLC) in Collective Perception (CP) — a technology that enables vehicles and infrastructure to share sensor information to help reduce traffic accidents. Recent advances in Vehicle-to-Everything (V2X) communication have spurred growing research interest in CP. However, in regions such as the United States and Japan, only 30 MHz of radio spectrum is allocated for V2X, which is insufficient to effectively support CP.

In this paper, we propose integrating VLC into V2X systems to enhance CP, complementing the existing 5.9 GHz band for V2X communications. VLC can coexist with wireless systems that use radio waves, providing an additional optical channel for data exchange. To the best of the authors’ knowledge, this is the first study to investigate VLC for CP.

We evaluate the feasibility of VLC-based CP through three experiments. First, we measured the application-level delay of a VLC-based CP system in a stationary indoor environment. Next, we evaluated its communication range in a stationary outdoor setting. Finally, to assess robustness under realistic conditions, we conducted driving experiments at vehicle speeds up to 90 km/h. The results demonstrate that VLC-based CP is feasible and could serve as a promising solution to spectrum scarcity in the 5.9 GHz band for future V2X communications.
%

\end{abstract}


\section{Introduction}
Collective Perception (CP) \cite{b_ETSI2019} technology can play a critical role in enhancing advanced driver-assistance systems (ADAS) and autonomous driving (AD) by enabling vehicles and infrastructure to exchange sensor information. This capability significantly extends the field of view beyond individual sensors, reducing blind spots and improving situational awareness \cite{b_CP}. CP leverages vehicle-to-everything (V2X) communication, including vehicle-to-vehicle (V2V) and vehicle-to-infrastructure (V2I)\cite{b_ETSI2019}, to mitigate inherent sensor limitations such as restricted fields of view and obstructions. However, the effectiveness of CP is challenged by limited frequency allocations in the widely utilized 5.9 GHz band, congestion in densely populated areas, and interference with other applications such as Wi-Fi \cite{b_ITSamerica,b_JK}.

Addressing these challenges requires examining complementary communication methods or additional frameworks that can be integrated with existing systems to help resolve these issues.
Visible Light Communication (VLC), a technology that transmits data via rapidly flashing LEDs imperceptible to human vision \cite{b_VLC_yamazato2, b_VLC_kinoshita}, offers a promising solution due to its directional properties and spatial separation capabilities. 
VLC inherently supports multicast transmission from headlights, taillights, traffic lights, signage, and Smartpoles, effectively aligning with CP requirements.

\begin{figure}[t!]
\centering 
\includegraphics[width=0.8\columnwidth]{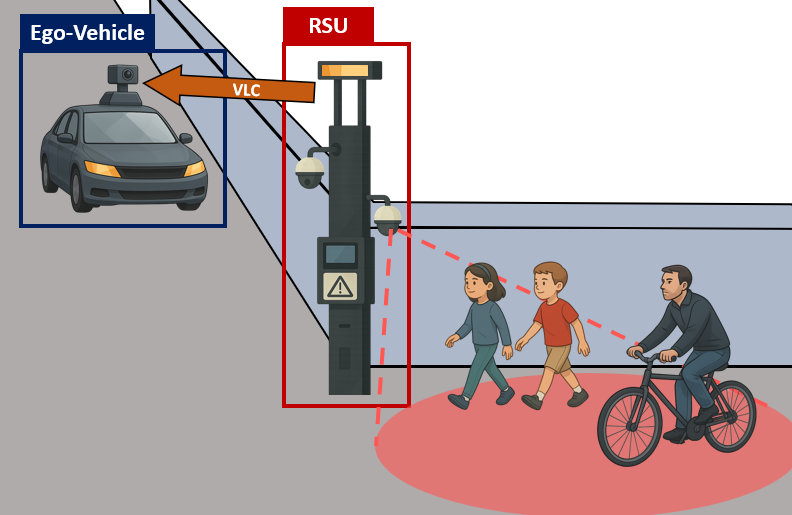}
\caption{A VLC-based CP system considered in this paper. In this figure, the Ego-Vehicle is attempting to pass through the intersection. The bike and the children are entering the intersection from a direction that is hidden by the wall from the Ego-Vehicle’s view. On the other hand, the RSU detects both the bike and the children entering the intersection and sends this information to the Ego-Vehicle via a VLC signal. Thus, the Ego-Vehicle can also recognize the objects behind the wall, which are normally unseen, and can safely pass through the intersection.}
\label{fig:VLC_based_CP}
\vspace*{-3mm}
\end{figure}

Figure \ref{fig:VLC_based_CP} illustrates the VLC-based CP system at an intersection. In this system, the Road-Side Unit (RSU) provides vehicles approaching or passing through the intersection with information about the positions and speeds of nearby objects. This helps vehicles navigate intersections with limited visibility.

This study proposes a novel VLC-based CP system that can complement traditional radio-based V2X systems, aiming to alleviate their inherent bandwidth limitations and congestion issues. By using high-speed cameras and LED transmitters integrated into the traffic infrastructure, the proposed system aims to minimize the data volume transmitted and optimize communication latency. Experimental evaluations demonstrate that VLC-based CP effectively meets the stringent timing constraints set by the European Telecommunications Standards Institute (ETSI) and achieves reliable communication performance even in mobile scenarios, thus indicating its potential for enhancing safety and efficiency in intelligent transport systems (ITS).

The main contributions of this study are:
\begin{itemize}
    \item Proposing and developing a novel VLC-based CP system to address bandwidth constraints and congestion issues inherent in traditional radio-based CP systems.
    \item Demonstrating that VLC technology can effectively meet the time constraint and the data-rate requirement of 50 bytes for VLC-based CP, thereby enhancing the safety and performance of ADAS and AD.
    \item Providing experimental validation of the VLC-based CP system under both stationary and mobile scenarios, confirming its feasibility and reliability for real-world deployment.
\end{itemize}

The paper is organized as follows: 
Section II discusses trends in CPM and the allocation of 5.9 GHz V2X spectrum. 
Section III provides a survey on CPM in the context of VLC and clarifies that the data size of CPM in VLC-based CP is approximately 50 bytes.
Section IV provides a detailed overview of the proposed VLC-based CP system. Section V discusses the experimental setup and presents the results of both stationary and mobile communication scenarios. Finally, Section VI concludes the paper by summarizing the key findings.

\section{Collective Perception (CP) and 5.9 GHz V2X Spectrum Allocation}
CP is a technology that enables vehicles and infrastructure to share and fuse sensor information \cite{b_ETSI2019}. CP aims to overcome the limitations of individual sensors (e.g., field-of-view limitations, shielding, bad weather) by allowing each vehicle to share information about its surroundings through vehicle-to-vehicle (V2V) and vehicle-to-infrastructure (V2I) communications \cite{b_CP}. This is expected to improve recognition accuracy and reduce reaction time \cite{b_CP_matsushita}.

On the other hand, not enough frequency bands have been allocated for V2X communication.
For example, Table \ref{table:v2x_spectrum_two_column} summarizes the frequency allocation trend in the 5.9 GHz band for V2X communications in the United States, Europe, Japan, and China. As summarized in the table, each country allocates several tens of MHz (equivalent to multiple channels) of the 5.9 GHz band for V2X communication.
However, while the 30 MHz bandwidth can support the transmission of Basic Safety Messages, including vehicle position, speed, and acceleration, as noted by ITS America \cite{b_ITSamerica}, a wider bandwidth is necessary to support more advanced functions such as CP \cite{b_JK}.
This need is particularly pressing in urban areas or locations with heavy traffic, where simultaneous communication among many vehicles can lead to communication congestion. 

One way to achieve CP at 30 MHz is to minimize the amount of Collective Perception Message (CPM) shared, which contains information about sensed objects, such as position and speed. In Europe, Decentralized Congestion Control (DCC) has already been standardized. However, methods that reduce transmission frequency, such as DCC, can introduce communication delays and information loss, which in turn may impact safety and reliability \cite{b_DCC_ETSI}. Additionally, DCC requires optimal control that adapts to dynamic traffic conditions and varying communication loads. Such adaptation necessitates further advancements in algorithms and real-time processing capabilities \cite{b_DCC_CHALLENGE}.

Therefore, this study proposes a novel approach for CPM transmission and reception using VLC, referred to as VLC-based CP.

\begin{table}[t!]
\vspace*{-3mm}
\centering
\caption{5.9 GHz V2X Spectrum Allocation by Region with References}
\label{5.9G-V2X-Spectrum-allocation}
\begin{tabular}{|l|p{6cm}|}
\hline
\textbf{Region} & \textbf{Details} \\
\hline
United States \cite{b_FCC_5_9GHz}& 
\textbf{Allocated Band:} 5.895–5.925 GHz (30 MHz) \newline
\textbf{Technology:} C-V2X \newline
\textbf{Status:} 
Final FCC rules adopted in Nov. 2024. 
DSRC sunset planned. \newline
Lower 45 MHz reallocated to Wi-Fi (unlicensed). \\
\hline
Europe  \cite{b_EC_5_9GHz}& 
\textbf{Allocated Band:} 5.855–5.925 GHz
(This 70 MHz also includes a band reserved for coexistence with urban rail ITS. Service Channel 1 (10 MHz) is designated for the transmission of CPM.) \newline
\textbf{Technology:} Technology: ITS-G5 \newline
\textbf{Status:} operational specifications defined in the C-Roads Roadside ITS-G5 System Profile v2.3.0 (2025).
\\
\hline
Japan \cite{b_japan_5_9GHz} & 
\textbf{Allocated Band:} 
5.895–5.925 GHz (30 MHz) is under consideration.\newline
\textbf{Technology:} Under consideration \newline
\textbf{Status:}Preparation of the 5.9 GHz band for V2X is ongoing, including coexistence studies and early trials.
Frequency allocation is targeted by FY2026.\\
\hline
China \cite{b_china_5_9GHz}& 
\textbf{Allocated Band:} 5.905–5.925 GHz (20 MHz) \newline
\textbf{Technology:} LTE-V2X \newline
\textbf{Status:} MIIT officially allocated the 5905–5925 MHz spectrum before 2020. Widespread field trials and commercial C-V2X deployment are ongoing; future bandwidth expansion is planned.\\
\hline
\end{tabular}
\label{table:v2x_spectrum_two_column}
\end{table}

\section{CPM in VLC system}
\subsection{Visible Light Communication (VLC)}
VLC is a communication technology that transmits data by rapidly flashing light sources, such as LEDs, at a speed imperceptible to the human eye \cite{b_VLC_yamazato2}. The receiving device, such as a camera or photodiode, detects and decodes the Visible Light signals to extract the transmitted information \cite{b_VLC}.

Various studies have explored the application of VLC technologies in the field of Intelligent Transport Systems (ITS) \cite{b_VLC_ITS2,b_VLC_ITS3,b_VLC_ITS4}. When applying VLC to ITS, cameras are widely employed as receivers due to their capability to capture signals over a wide field of view. On the transmitters' side, vehicle headlights, taillights, and traffic signals are typically considered. 

\begin{figure}
    \centering
    \includegraphics[width=0.8\columnwidth]{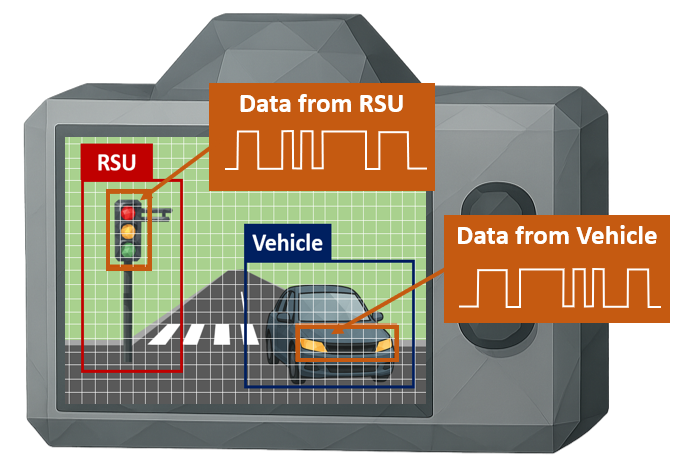}
    \caption{An example of camera-based VLC. The VL signals within the image data can be easily separated, while visual information is also obtained.\cite{b_VLC_yamazato2}}
    \label{fig:VLC_view}
\vspace*{-6mm}
\end{figure}

Figure \ref{fig:VLC_view} presents an image captured by a camera within a VLC system designed for ITS applications. Since light has directional properties, signals transmitted from separate ITS Stations can be readily distinguished.
Moreover, because VLC enables demodulation from visual information, the receiver can clearly recognize the type and location of each ITS Station, as illustrated in the figure. This means that even when multiple ITS stations are within line of sight (LoS), fast synchronization can be achieved. Although the use of VLC is subject to the strict limitation of requiring LoS, this constraint physically restricts the communication range, making interception or tampering from outside difficult. This provides a security advantage.

\subsection{Survey on CPM for VLC}
In scenarios illustrated in Figs. \ref{fig:VLC_based_CP} and \ref{fig:VLC_view}, low latency is crucial, whereas a small amount of data is generally sufficient \cite{b_SIP2020}. To discuss data volume more concretely, we conducted a comprehensive survey on CPM. The survey revealed that the time constraints, data capacity, and data content of CPM, which are established under the assumption of transmission solely via radio waves, are not well optimized for VLC-based CP. This is envisioned to play a complementary role alongside existing systems. We found that transmitting smaller data volumes at higher frequencies would more effectively utilize the advantages of VLC \cite{CPM_VLC}.

According to ETSI, CP via radio wave communication should be executed within 100 to 1000 ms and requires a data size of approximately 200 bytes \cite{b_ETSI2019}. Regarding time constraints, the upper limit defines the validity period of the information. If this duration (1000 ms) is exceeded, the state of the sensed object will no longer remain consistent between the moment it was sensed and the moment it is shared with the surrounding ITS stations.
This constraint would similarly apply to VLC-based CP, as it is independent of the communication medium. 

On the other hand, the lower limit was established to mitigate radio wave congestion, which can be ignored in VLC-based CP as it does not use frequency resources. The practical lower limit for VLC remains under investigation. However, future implementations transmitting only essential information in under 100 ms would likely yield superior performance in VLC-based CP.

The data rate proposed by ETSI is considered excessive, and various research is being conducted to reduce data transmission and frequency \cite{b_CPM_mitigation1,b_CPM_mitigation2}. Furthermore, since the proposed VLC-based CP employs a different communication medium, a reevaluation of the data content structure is also required. 

Table \ref{tab:vlc-cpm} illustrates the structural breakdown of CPM by ETSI \cite{b_CPM_structure}. CPM consists of HeaderInfo, Payload, and TrailerInfo. The HeaderInfo includes the information of the psid, which carries the ITS-AID of the message, and the generationTime. The Payload contains identification information (ITS PDU Header), reference information required for correctly interpreting the message (Management Container), transmitter information (Originating Vehicle/RSU Container), perceived object data (PerceivedObject Container), and sensor information (SensorInformation Container). The TrailerInfo contains the cryptographic material that ensures the integrity and authenticity of the message. In the ETSI/IEEE security framework, this component typically carries an ECDSA signature, which is the mandated baseline scheme for ITS messages. The signature allows receivers to verify that the payload has not been altered and that it originates from a legitimate sender.

From Table \ref{tab:vlc-cpm}, it can be observed that the CPM proposed by ETSI requires approximately 200 bytes even to transmit the information of a single object.
Based on the above survey, this study positions VLC-based CP as a complementary approach to existing methods, with the expectation that it can contribute to alleviating spectrum scarcity and enhancing safety. Accordingly, the design policy is to transmit low-capacity data at high frequency. Specifically, the transmitted information is limited to the minimum set directly related to accident reduction, such as the object’s position, speed, and timestamp.
Regarding security, since ETSI mandates the use of ECDSA and VLC-based CP assumes high-frequency transmissions, we adopt a structure where ECDSA is transmitted separately at a lower frequency, while each message is accompanied by a Message Authentication Code (MAC).

In this study, we assume that a single VLC-based CPM carries information on two objects, and the discussion is based on a 50-byte transmission.
%
\begin{table*}[ht]
\centering
\caption{Comparison of CPM structures: ETSI standard vs. proposed VLC-based structure.}
\label{tab:vlc-cpm}
\renewcommand{\arraystretch}{1.1}
\setlength{\tabcolsep}{3pt}

\newcolumntype{L}[1]{>{\raggedright\arraybackslash}p{#1}}
\newcolumntype{C}[1]{>{\centering\arraybackslash}p{#1}}

\begin{tabular}{@{}L{2.2cm} L{3.5cm} C{1.3cm} @{\hskip 10pt}
                L{2.2cm} L{3.5cm} C{1.3cm}@{}}
\toprule
\multicolumn{3}{c}{\textbf{CPM structure as specified by ETSI}} &
\multicolumn{3}{c}{\textbf{Proposed VLC-based CPM structure}} \\
\cmidrule(lr){1-3} \cmidrule(lr){4-6}
\textbf{Component} & \textbf{Description} & \textbf{Size [bits]} &
\textbf{Component} & \textbf{Description} & \textbf{Size [bits]} \\ 
\cmidrule(lr){1-3} \cmidrule(lr){4-6}
\multicolumn{3}{l}{\textit{HeaderInfo}} & \multicolumn{3}{l}{\textit{HeaderInfo}} \\
psid & Provider Service Identifier & 16 &
psid & Provider Service Identifier & 16 \\
generationTime & Message generation time & 64 &
generationTime & Message generation time & 64 \\
\cmidrule(lr){1-3} \cmidrule(lr){4-6}
\multicolumn{3}{l}{\textit{Payload}} & \multicolumn{3}{l}{\textit{Payload}} \\
ITS PDU header & Protocol Version, MessageID, StationID& -- &ITS PDU header & Protocol Version, MessageID, StationID & 8+8+32\\
Management Container & Common management information including Originating Vehicle/RSU & 968 (including IST PDU header)&Management Container & Reference Position and Reference Time only & 64+16 \\
Perceived Object Data & Object-level information & 280$\times n$ & Perceived Object Data & Coordinate \textit{x,y} (16 each), Velocity \textit{v$_x$,v$_y$} (15 each), $\Delta t$ (12) & 74$\times n$ \\
Sensor Information & Sensor type, pose, and FOV & 280$\times m$  \\
\cmidrule(lr){1-3} \cmidrule(lr){4-6}
\multicolumn{3}{l}{\textit{TrailerInfo}} & \multicolumn{3}{l}{\textit{TrailerInfo}} \\
ECDSA & Digital signature (unspecified key length) & 512 &
MAC (64-bit) & Lightweight authentication & 64 \\
 & & &
ECDSA & Sent separately (low frequency) & -- \\
\cmidrule(lr){1-3} \cmidrule(lr){4-6}
\textbf{Total} &  & \hspace*{-10mm} \textbf{1560 + 280($m$+$n$)} &
\textbf{Total} &  & \hspace*{-6mm} \textbf{272 + 74$n$} \\
\bottomrule
\end{tabular}

\vspace{2pt}
\footnotesize
$n$: number of perceived objects; $m$: number of sensors.  
\textbf{VLC advantage:} Using cameras, VLC conveys sensor pose and field of view through visual data. The directionality of light helps receivers distinguish signals from different ITS stations, while visual demodulation reveals type and location, reducing CPM size.
\vspace*{-5mm}
\end{table*}

\section{SYSTEM OVERVIEW}

\begin{figure}
    \centering
    \includegraphics[width=1.0\columnwidth]{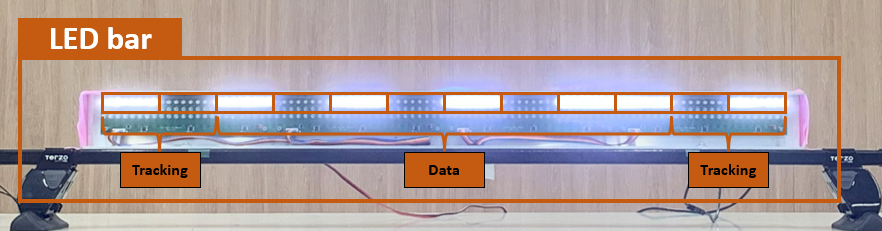}
    \caption{ LED bar transmitting 1 symbol using 8 LEDs. }
    \label{fig:ledbar}
\vspace*{-3mm}
\end{figure}

We use the LED bar transmitters illustrated in Fig. \ref{fig:ledbar}, which are composed of 1×96 LEDs, where the brightness of each LED is controlled by two voltage levels (ON and OFF).
On–Off-Keying (OOK) is adopted because, at long communication distances, it is difficult to reliably distinguish fine brightness differences, making multilevel modulation impractical.
This allows the LED Bar to transmit up to 96 bits simultaneously. In this study, to enable long-distance communication, we consider eight LEDs as a single symbol transmission block, allowing for the simultaneous transmission of 12 bits. Moreover, the two blocks at both ends are used to track the LED bar's position; only 8 blocks (8 bits) can be transmitted at one time.
LED detection is performed by frame differencing between adjacent images, allowing only LEDs blinking at the intended frequency to be extracted. The LED bar is then tracked by updating the region of interest (ROI) based on the tracking signals located at both ends of the detected LED bar.

The receiver uses a high-speed camera to capture images of the LED Bar and receive the visible light signals.
The images are captured in grayscale, with each pixel assigned an integer value between 0 and 255 (8 bits). During signal reception, the LED Bar is detected within the image, and the symbols represented by each LED block are determined using thresholding.

This system conducts unidirectional communication from the RSU to the Ego-Vehicle, which is equipped with a camera and an LED bar.
In this paper, we define the communication delay as the time interval from when the RSU starts transmitting data until the Ego-Vehicle completes demodulation.
\begin{figure}
    \centering
    \includegraphics[width=1.0\columnwidth]{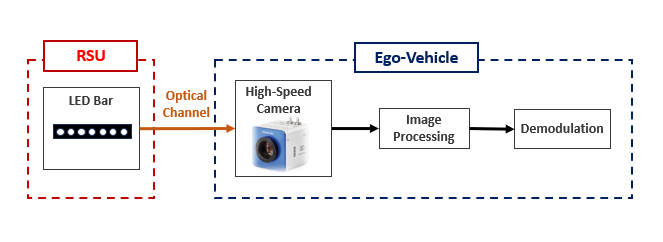}
    \vspace*{-12mm}
    \caption{Block diagram of the VLC-based CP system utilizing LED Bar and high-speed camera. }
    \label{fig:sysmodel}
\end{figure}


\section{Experiments}
This section presents experiments we conducted to evaluate VLC-based CP. First, we experimentally verified the delay required to execute VLC-based CP when the RSU and Ego-Vehicle were stationary. Furthermore, we investigated the communication range performance and the performance under driving conditions in an outdoor environment. Outdoor experiments were conducted exclusively during the daytime under clear weather conditions, and the only external environmental factor was sunlight.

\subsection{Application-level latency of VLC-Based CP }
\subsubsection{Overview of the experiment}
This experiment aims to measure the delay during 50-byte transmission in VLC-based CP. The experiment was conducted indoors, with both the RSU and the Ego-Vehicle kept stationary. The distance between the two was 5 m.

Since CP contributes to safety improvements, it is necessary to consider delays at the application level \cite{b_CPM_delay}.
In camera-based VLC, processing is performed on the received images, which makes the camera's frame rate and image processing speed key factors in this context. Therefore, as shown in Table \ref{table:ex_indoor}, we measured the delay while varying the frame rate from 100 fps to 1000 fps.
If the image processing finishes during the camera’s frame interval, the communication delay can be expressed in a very simple form, as shown in (\ref{eq:latency}).

\begin{equation}
    \tau =  
    \left(\frac{2N_{\text{frame}}}{\text{fps}}\right) 
\label{eq:latency}
\end{equation}
Here, $N_{\text{frame}}$ refers to the number of transmitting frames of the LED bar.

In the experiment, we used the system model shown in Fig. \ref{fig:sysmodel}. 
The application-level delay of the VLC-based CP is calculated by measuring the time difference between the transmission timestamp recorded when data is written to the LED bar on the RSU side and the reception timestamp when the signal is demodulated on the Ego-Vehicle side.

\begin{figure}
    \centering
    \includegraphics[width=0.8\columnwidth]{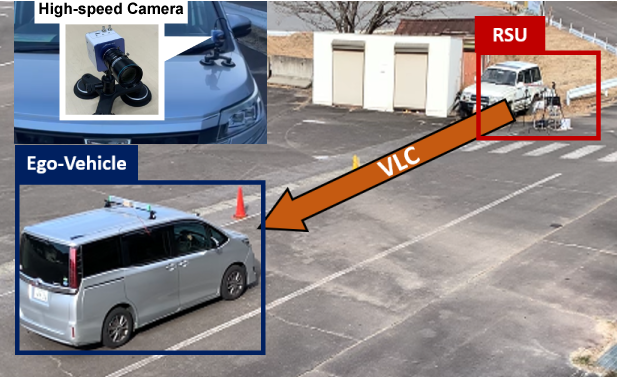}
    \caption{ Experiment setup of our proposed system in both a stationary outdoor setting and driving environment.}
    \label{fig:ex_outdoor}
\end{figure}

\begin{table}
\vspace*{-3mm}
    \centering
        \caption{Specifications for experiment in the stationary scenario}
\begin{tabular}{|l|c|c|}
        \hline
        Transmitter       & LED Bar  \\
        Sending speed     & 50, 150, 100, 125, 200, 250, 400, 500 Hz \\
        Data per frame    & 1 byte/frame \\
        $N_{\text{frame}}$        & 53 frame \\
        \hline
        Reciever          & Photron INFINICAM \\
        Camera resolution & 600 × 320 \\
        Shutter Speed     & 1 / 2000 \\
        Frame rate        & 100, 125, 200, 250, 400, 500, 800, 1000 fps \\
        Aperture          & F16 \\
        Focal length      & 12.5 mm \\
        \hline
        Enviroment        & Indoor / Stationary\\
        Distance          & 5 m \\
        \hline
    \end{tabular}
    \label{table:ex_indoor}
\vspace*{-3mm}
\end{table}

\subsubsection{Results of delay measurement}

\begin{figure}[ht]
    \centering
    \includegraphics[width=1.0\columnwidth]{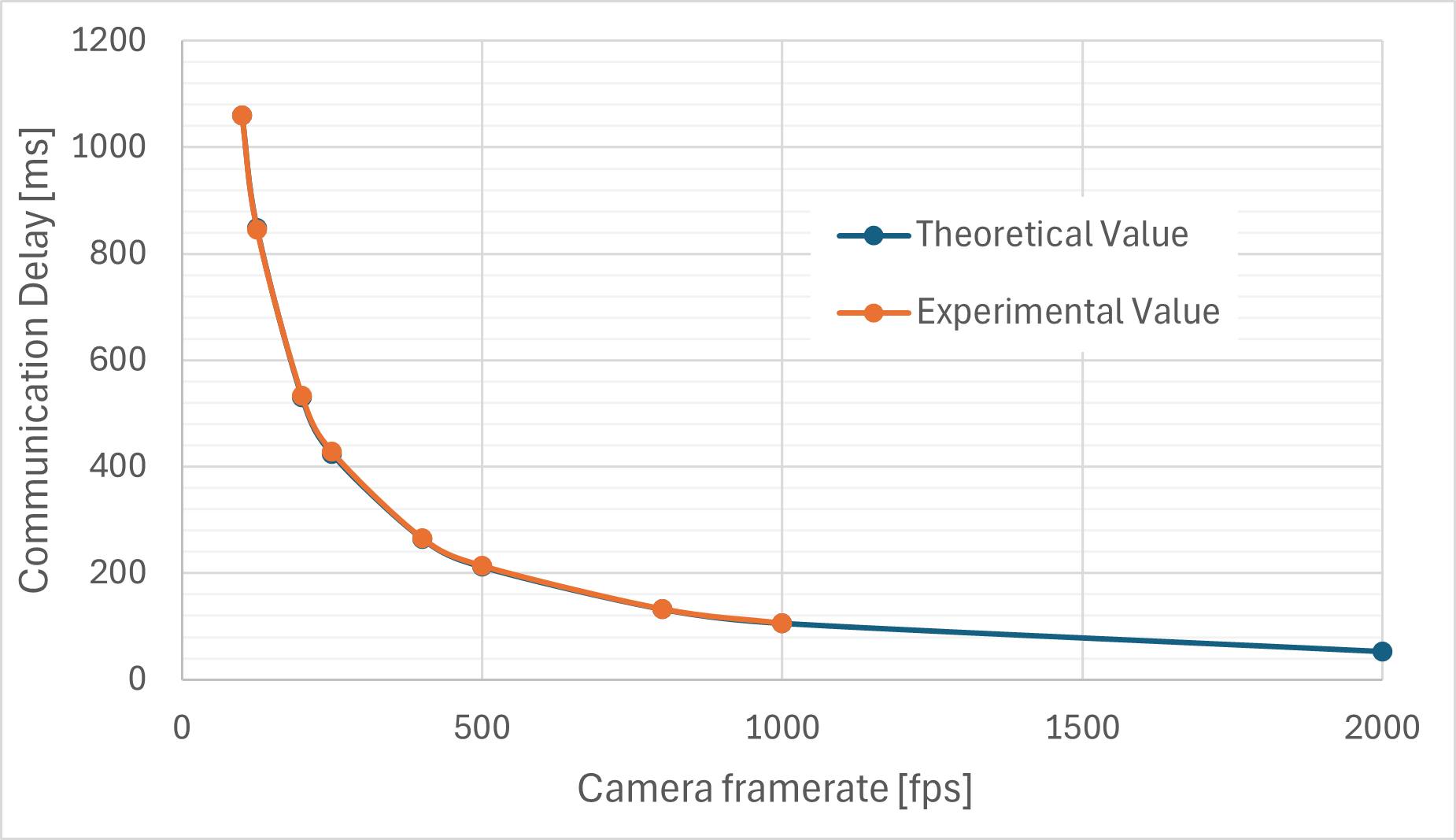}
    \caption{Camera fps versus Round Trip Time, which is equivalent to the application-level latency required for VLC-based Cooperative Perception.}
    \label{fig:result_delay}
\end{figure}
The experimental result is shown in Fig. \ref{fig:result_delay}. For reference, error-free communication was achieved in this experiment. 
Acquiring an image takes at least as long as the frame rate, linking latency to the frame rate. As the frame rate increases, latency decreases until a threshold is reached, after which processing time dominates, and further increases in frame rate do not reduce latency.
The latency shown in Fig. \ref{fig:result_delay} indicates the time for VLC-based CP, with theoretical and experimental values aligning well.

ETSI specifications state that CPMs should be transmitted within 100 to 1,000 ms. In the VLC-based CP system at a frame rate of 1,000 fps, the entire process—from transmission to demodulation—can be completed in about 100 ms.

\subsection{Communication Range Performance}
\subsubsection{Overview of the experiment}
To verify that the proposed system can be applied to actual driving experiments, we investigated the relationship between the communication range and the bit error rate (BER) through experiments.
Table \ref{table:ex_outdoor} summarizes the experimental parameters, and Fig. \ref{fig:ex_outdoor} shows the experimental setup.
In this experiment, both the RSU and the Ego-Vehicle were stationary, and the Ego-Vehicle demodulated 200 packets of 50 bytes each. The BER was then calculated based on the received data.
\begin{table}
\vspace*{-3mm}
    \centering
        \caption{Specifications for experiment in the mobile scenario}
\begin{tabular}{|l|c|}
        \hline
        Transmitter       & LED Bar  \\
        Sending speed     & 500 Hz \\
        Data per frame    & 1 byte/frame \\
 $N_{\text{frame}}$        & 53 frame \\
        \hline
        Reciever          & Photron INFINICAM \\
        Camera resolution & 600 × 320 \\
        Shutter Speed     & 1 / 2000 \\
        Frame rate        & 1000 fps \\
        Aperture          & F16 \\
        Focal length      & 100 mm \\
        \hline
        \multicolumn{2}{|c|}{{Communication range experiment.}} \\
        \hline
        Enviroment        & Outdoor, Daytime / Stationary\\
        Distance          & 10 $\sim$ 150 m  \\
        \hline
        \multicolumn{2}{|c|}{{Vehicle driving experiment.}} \\
        \hline
        Enviroment        & Outdoor, Daytime / Driving\\
        Distance          & 100 $\sim$ 120 m \\
        Driving Speed     & 20, 40, 60, 90 km/h \\
        \hline
    \end{tabular}
    \label{table:ex_outdoor}
\vspace*{-3mm}
\end{table}

\subsubsection{Results of communication range experiment}
As indicated in Table IV, the BER was on the order of \(10^{-4} \) for communication distances ranging from 75 m to 160 m.
Also, this work employed OOK modulation exclusively, and the demodulation performance could be improved by applying error-correction coding, such as Polar codes.

This experimental result demonstrates that communication beyond 100 m is sufficiently achievable using the proposed system.

\begin{table}
\vspace*{-5mm}
    \centering
        \caption{BER vs. Communication Range}
\begin{tabular}{|c|c|}
        \hline
        Communication Range [m]     & BER  \\
        \hline
        75  & \( 2 \times 10^{-4} \) \\
        100 & \( 1.38 \times 10^{-4} \) \\
        120 & Error Free \\
        140 & Error Free \\
        160 & \( 8 \times 10^{-4} \) \\
        \hline
    \end{tabular}
    \label{table:ex_outdoor_BER}
\vspace*{-3mm}
\end{table}

\subsection{Trials of mobile communication assuming actual use cases }
\subsubsection{Overview of the experiment}
We conducted a trial experiment to evaluate VLC-based CP. During the experiment, as shown in Fig. \ref{fig:ex_outdoor}, the car was driven towards the LED transmitter from a distance of 200 m. The vehicle speeds were set at 20, 40, 60, and 90 km/h as shown in Table~\ref{table:ex_outdoor_BER}. Additionally, the communication between the RSU and the Ego-Vehicle was established at a distance of 100 to 120 m, determined by the available distance for safe acceleration and deceleration on the test track.
The driving experiments were conducted four times at each speed.
In each run, 50 bytes of data were transmitted from the RSU to the Ego-Vehicle, and the BER was calculated based on the received data.

\subsubsection{Results of driving trials}
\begin{figure}
    \centering
    \includegraphics[width=1.0\columnwidth]{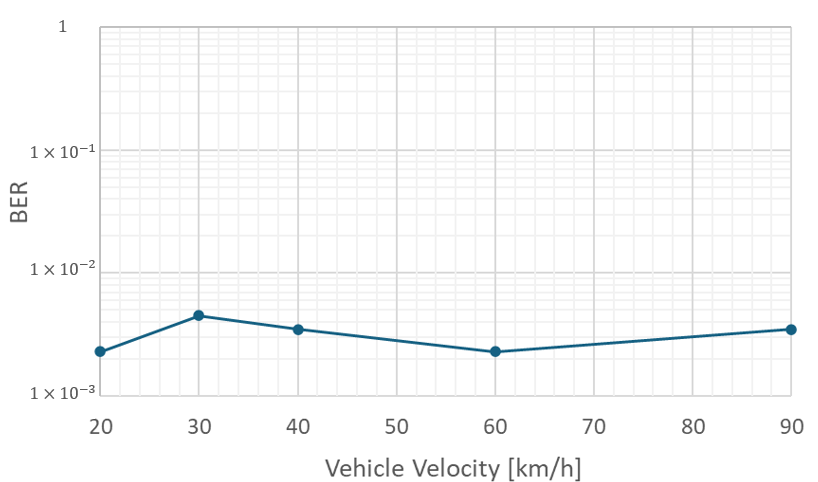}
    \caption{BER measurement result in driving experiments at vehicle speeds up to 90 km/h, with the communication distance up to 120 m.}
    \label{fig:result_driving}
\end{figure}

The BER shown in Fig.~\ref{fig:result_driving} represents the upper bound of the 95\% confidence interval, considering the variation in the measurement results. 
The values were on the order of $10^{-3}$ and were not affected by the driving speed, meaning that error-free communication is possible by applying a powerful error-correcting code.

\section{Conclusion}
Conventional CP implementations that rely solely on radio-based V2X communication face challenges such as spectrum congestion.
To address this issue, we proposed a VLC-based CP system using an LED transmitter and a  camera receiver.

We analyzed existing CP message formats and proposed a message structure suitable for camera-based CP. A prototype system was implemented and experimentally evaluated. The results confirmed that a 50-byte payload can be transmitted within 100 ms in terms of latency performance.Furthermore, driving experiments demonstrated robust communication at vehicle speeds of up to 90 km/h and distances exceeding 100 m.

Future work includes the evaluation of communication performance in multi-node environments and under Non-Line-of-Sight (NLoS) conditions.



\end{document}